# GWAPP: A Web Application for Genome-wide Association Mapping in *A. thaliana*.

Ümit Seren[a,1], Bjarni J. Vilhjálmsson[a,b,1,2], Matthew W. Horton[a,c], Dazhe Meng[a,b], Petar Forai[a], Yu S. Huang[d], Quan Long[a], Vincent Segura[e], Magnus Nordborg[a,b]

a.  Gregor Mendel Institute, Austrian Academy of Sciences, Vienna, Austria
b.  Molecular and Computational Biology, University of Southern California, CA, USA
c.  Department of Ecology and Evolution, University of Chicago, Chicago, IL, USA.
d.  Center for Neurobehavioral Genetics, Semel Institute, University of California Los Angeles, CA, USA
e.  INRA, UR0588, F-45075 Orléans, France.

[1] These authors contributed equally.

[2] Current address: Department of Epidemiology, Harvard School of Public Health, Harvard University, Boston, MA, USA.

## Abstract

*Arabidopsis thaliana* is an important model organism for understanding the genetics and molecular biology of plants. Its highly selfing nature, small size, short generation time, small genome size, and wide geographic distribution, make it an ideal model organism for understanding natural variation. Genome-wide association studies (GWAS) have proven a useful technique for identifying genetic loci responsible for natural variation in *A. thaliana*. Previously genotyped accessions (natural inbred lines) can be grown in replicate under different conditions, and phenotyped for different traits. These important features greatly simplify association mapping of traits and allow for systematic dissection of the genetics of natural variation by the entire *Arabidopsis* community. To facilitate this, we present GWAPP, an interactive web-based application for conducting

GWAS in *A. thaliana*. Using an efficient implementation of a linear mixed model, traits measured for a subset of 1386 publicly available ecotypes can be uploaded and mapped with a mixed model and other methods in just a couple of minutes. GWAPP features an extensive, interactive, and a user-friendly interface that includes interactive Manhattan plots and LD plots. It also facilitates exploratory data analysis by implementing features such as the inclusion of candidate SNPs in the model as cofactors.

## Introduction

Genome-wide association studies (GWAS) are rapidly becoming the dominant paradigm for investigating the genetics of natural phenotypic variation. Although GWAS have primarily been used for human diseases, they have also been successful in mapping causal variants in many other organisms, including *A. thaliana*, which is an ideal organism for such studies. In particular, the ready availability of diverse inbred lines that have already been genotyped means that it is possible for anyone to carry out GWAS by simply ordering and phenotyping these lines (Atwell et al., 2010; Todesco et al., 2010; Baxter et al., 2010; Horton et al., 2012). The only remaining obstacle is the statistical analysis. *A. thaliana* generally displays strong and complex population structure, mainly due to isolation by distance (Platt *et al.*, 2010), and this must unequivocally be taken into account in any GWAS (Aranzana et al., 2005; Atwell et al., 2010). The only statistical method that appears to be effective for this purpose in *A. thaliana* is a mixed model that takes population structure into account using a genetic relatedness matrix (Yu *et al.*, 2006; Zhao *et al.*, 2007). Software that implements these models exists (Bradbury *et al.*, 2007; Kang *et al.* 2010; Zhang *et al.*, 2010; Lippert *et al.* 2011; Lipka *et al.*, 2011; Zhou and Stephens, 2012; Svishcheva *et al.* 2012), but requires the user to provide both the genotype and phenotype data, as well as filtering and ordering the data appropriately. In addition, they provide little or help in analyzing the results. Some of these concerns were recently addressed in *Matapax* (Childs *et al.*, 2012), a web-based pipeline for conducting GWAS in *A. thaliana*, which includes some interactive features but still requires the user to wait hours for the results.

Here we present GWAPP, a user-friendly and interactive web application for GWAS in *A. thaliana*. GWAPP places a strong emphasis on informative and efficient visualisation tools for interpreting the GWAS results and provides interactive features that allow for hands-on in-depth analysis. Using efficient implementations of both a Wilcoxon rank-sum test and an approximate mixed model (Zhang *et al.*, 2010; Kang *et al.* 2010), the mapping is performed on-the-fly, with genome-wide scans for ~206,000 SNPs and 1386 individuals completed in minutes. GWAPP enables the user to view, select subsets, and choose an appropriate transformation before carrying out the GWAS. It allows the inclusion of SNPs as cofactors in the model in an interactive manner, and provides guidelines for how to do this. With interactive Manhattan plots of association p-values along the chromosomes, GWAPP allows for a quick summary of the results, as well as visualisations of both genome-wide and local linkage disequilibrium patterns. By zooming in on certain regions of interest, down to gene-level, the p-values are displayed together with the gene models and their annotation in fast interactive plots. We also display population genetic statistics, including selection scores and recombination rate estimates (Horton *et al.*, 2012). GWAPP can be accessed at: **http://gwas.gmi.oeaw.ac.at**, all code is public and can be obtained at **http://code.google.com/p/gwas-web-app/**.

## Results

### User Interface

GWAPP consists of a web-front-end with a graphical user interface, and a back-end that handles the data and performs the mapping. The main menu in the top section contains five entries that allow access to different functions of the web-front-end. *HOME* is the landing page and provides general information about GWAPP and a quick tutorial. A more detailed tutorial and description of the functionality can be found under the *HELP* tab. The *ACCESSIONS* section displays a list of the 1386 publicly available *A. thaliana* accessions for which GWAPP provides genotype data. This page also displays information about the geographic distribution of the data set and the location of each accession. In the *UPLOAD PHENOTYPE* section, phenotypes can be uploaded. The server supports multiple phenotypes, and these are stored, together with any analysis results, on the server and tied to a unique dataset-key and a cookie on the user's client computer. This allows the user to continue the analysis from where he left off, without having to redo any previous analysis steps, from a different computer. The *ANALYSIS* section is the most important part of the web-interface, where users can view the uploaded phenotypes, create datasets, apply transformations, run GWAS and analyze the results. We will discuss these features further in the following sections.

## *ANALYSIS* page

Once a phenotype file has been uploaded, the user can verify and view the phenotype(s) on the *ANALYSIS* page. The page is split into two sections: (1) a hierarchical tree on the left side in the 'Navigation' box allows quick access to different phenotypes that have been uploaded and four levels of information (Supplemental Figure 1); (2) the right section of the *ANALYSIS* page, which is used for displaying the main content. The components in the hierarchical tree reflect the stored phenotype data-structure. The four levels of information are: (a) the root level contains phenotype-information; (b) each phenotype contains one or more subsets of the data; (c) each dataset can have one or more transformations; (d) each transformation can contain one or more GWAS results.

**Phenotype view**

The phenotype view (Figure 1) is visible upon selecting a specific phenotype from the left navigation tree. Information is displayed in three related information panels. The top panel contains general information about the selected phenotype (Figure 1A). The center panel contains a list of all datasets, where a dataset is defined as a subset of lines with phenotype values, together with the geographic distribution of the phenotyped accessions (Figure 1B, 1C). By default every uploaded phenotype contains a 'Fullset' dataset that contains all the phenotype values available. The plot showing the geographical distribution is updated when a different dataset (subset) is chosen. In the bottom panels (Figure 1D), basic statistics are shown for all the datasets.

**Dataset view**

4— 

By choosing an existing dataset or by creating a new one the user is directed to the dataset view (Figure 2). This view consists of a list of the accessions and two geographical plots. Using the list of accessions, the user can edit or create new datasets/subsets that contain, for example, only accessions from a specific country or collection. Using the list and the geographical map, the user can exclude, or include, certain accessions from specific regions. As different advantageous alleles can be expected to arise in some local adaptation scenarios (Chan *et al.*, 2010), it may be beneficial for some traits to use regional datasets for mapping causal alleles.

**Transformation view**

Applying a transformation to the phenotype may result in more reliable results for parametric tests. The transformation aims to facilitate the process of selecting a reasonable transformation, allowing the user to instantly preview the resulting phenotypic distribution. The view consists of four panels (Figure 3). The center panel (Figure 3C) contains a phenotype-explorer component (Huang *et al.*, 2011), which, among other things, allows the user to plot phenotype values against latitude and longitude in a motion chart.

The transformations implemented include logarithmic, square root, and Box-Cox transformations (Box and Cox, 1964). The p-value for Shapiro-Wilks test for normality is reported in the histogram and may assist in choosing an appropriate transformation. However, we want to note that choosing an appropriate transformation in structured samples is not trivial since phenotypes are expected to have a multivariate distribution with non-zero correlations (Fisher, 1918). Since the phenotypes are not independent observations, their distribution may deviate from a bell shaped univariate Gaussian distribution, even if they follow a multivariate Gaussian distribution. After deciding on a transformation, a genome-wide association scan can be performed. In the current version, the user can choose between (1) a non-parametric Wilcoxon rank-sum test (Wilcoxon, 1945), (2) a simple linear regression (LM), and (3) an accelerated mixed model (AMM). AMM first performs a genome-wide scan using the approximate inference proposed by (Zhang *et al.*, 2010; Kang *et al.*, 2010), and then updates the smallest 100 p-values using an exact mixed model inference (Kang *et al.*, 2008). Both LM and AMM employ a parametric F-test to obtain the p-values. For examining p-value bias due to population stratification one can use the Kolmogorov-Smirnov statistic, the median p-value, as well as the QQ-plots (Atwell *et al.* 2010).

**Results view**

The results view has two main components that can be accessed via the *Plots* and *Statistics* tabs. Under the *Plots* tab an interactive Manhattan plot (a scatter-plot with the negative logarithm p-values for the SNP association plotted against the SNP positions) for all five chromosomes is shown. The Benjamini–Hochberg–Yekutieli multiple testing procedure (Benjamini and Yekutieli, 2001) was used to control the false discovery rate. Assuming arbitrary dependence between SNPs, the 5% FDR threshold is plotted as a dashed horizontal line. Moving the mouse over a specific point in the plot will display the position of the corresponding SNP and its p-value. The Manhattan plot supports zooming, which can be achieved by a 'click, hold and drag mouse' gesture which defines the area for the zoom action. If the zoom level is below a specific threshold (~ 1.5 Mb) a gene annotation view (GeneViewer), which we developed specifically for this application, is displayed (Figure 4). Moving the mouse over a point in the Manhattan plot will also display a vertical line in the gene annotation view. If the zoom level is below 150 Kbp a more detailed gene annotation view containing gene features, e.g., the coding sequence region (CDS) and the untranslated regions (UTRs) will be shown. Moving the mouse over a specific gene in the GeneViewer will display a pop-up with additional functional description for the gene. Clicking on the gene will direct the user to *The Arabidopsis Information Resource* (TAIR) webpage for the gene, containing more detailed information. Finally, the user can highlight specific genes using a gene-search field above the scatter-plot for the first chromosome.

When the zoom-level in the p-value plot is below ~1.5Mb, a statistics panel is displayed below the gene annotations. By default the statistics panel will show the gene density as a filled bar-chart, but other statistics can be selected by clicking on the gears icon. Currently, five other statistics can be chosen: (1) Fst (North - South) (Lewontin and Krakhauer, 1973); (2) CLR [(Nielsen *et al.*, 2005); (3) PHS (Toomaijan *et al.*, 2006); (4) recombination estimate (rho) (McVean *et al.*, 2004); (5) sequence similarity with *Arabidopsis lyrata* (Hu *et al.*, 2011). Four of these statistics, Fst, CLR, PHS, and the recombination rate estimate, were calculated using the Horton *et al.* dataset (Horton *et al.*, 2012), and may not always be representative for the subset being analyzed. In an attempt to address this issue the user can upload statistics provided in a CSV file with 3 columns (chromosome, position and value). This further enables the user to plot miscellaneous statistics, which he may have and find useful, underneath the Manhattan plots. All of the plotted statistics are binned, and displayed in a similar interactive chart as the p-values, which also allows for vertical and horizontal zooming. The region that the user has zoomed in on in the p-value chart is highlighted in yellow. The bin size used to show the statistics can be adjusted by changing the number in the white textbox in the lower left corner of the plot.

Users can also visualize the linkage disequilibrium (LD) structure. This can be done by clicking on any SNP and choosing from three different methods: (1) Show LD in this region, (2) Calculate exact LD in this region, (3) Highlight SNPs in LD for this SNP. The first two options ( "Show LD in this region" and "Calculate exact LD in this region") display a LD triangle plot below the gene annotation panel and color-code the SNPs in the Manhattan plot (Figure 5). The difference between the first two options are that the former only displays $r^2$ values for the visible SNPs and the latter will calculate and show the $r^2$ values of all SNPs regardless if they are displayed or not. Both options display pair-wise $r^2$ values of at most 500 SNPs (due to limitations regarding visualization and computational complexity). For the sake of visual clarity, only $r^2$ values above 0.3 are color-coded. Furthermore, selecting a SNP in the Manhattan plot will color-code all neighboring SNPs according to their $r^2$ value. At the same time all pair-wise $r^2$ values in the triangle plot will be highlighted (Supplementary Figure 2). Similarly, when a specific $r^2$ value in the triangle plot is selected, the corresponding pair of SNPs in the Manhattan plot and the triangle plot is highlighted with corresponding color-coding (Supplementary Figure 3). Lastly, the third option ("Highlight SNPs in LD for this SNP") will calculate genome-wide $r^2$ values between the selected SNP and all other displayed SNPs and color-code them in the Manhattan plot (Supplementary Figure 4).

When using AMM, SNPs can be included as cofactors in the mixed model by clicking on a specific SNP and choosing 'Run Conditional GWAS'. This allows the user to perform conditional analysis on both local and global scale. Including causal loci in the model has been shown to be beneficial for finding other causal markers in structured data (Segura *et al.*, 2012; Vilhjálmsson and Nordborg, 2012). The second tab of the results view contains statistical descriptors and plots, which are useful when comparing models with different SNPs included as cofactors. These include three different model selection criteria: (1) the Bayesian information criterion (BIC) (Schwarz, 1978), (2) the extended Bayesian information criterion (e-BIC) (Chen and Chen, 2008), (3) and the multiple Bonferroni criterion (Segura *et al.*, 2012). These model selection criteria can guide the user to select reasonable models (and cofactors) in the absence of other prior knowledge. AMM has the added advantage that it estimates the variance components, from which the overall narrow sense heritability estimates (pseudo-heritability) can be obtained. When cofactors are included and the analysis is rerun, these estimates are updated, providing an overview of how the phenotypic variance is partitioned among three categories: (1) the fixed effects, i.e., the variance explained by the SNP cofactors; (2) the random genetic term, which estimates the amount of unexplained variance attributable to genetics; (3) the random error term, which is the fraction of variance attributed to random noise. These statistics provide a rough estimate of whether, and to what degree further genetic effects can be detected. Hence, if the remaining genetic fraction of phenotypic variance is small, there may not be much reason for including more cofactors in the model.

## A GWAPP analysis example for flowering time

To demonstrate how GWAPP can be used to make real biological discoveries we use a flowering time dataset published by (Li *et al.*, 2010). We focus on flowering time measured in 479 plants grown in growth chambers set to simulate Swedish spring conditions (Li *et al.*, 2010). Flowering time in *A. thaliana* has been extensively studied with both linkage mapping (Salomé *et al.*, 2011) and GWAS (Atwell *et al.*, 2010; Brachi *et al.*, 2010; Li *et al.*, 2010). Furthermore, several genes have been shown to harbor genetic variants that affect flowering time, including *FLOWERING LOCUS C* (*FLC*) (Michaels and Amasino, 1999), and *FRIGIDA* (*FRI*) (Johanson *et al.*, 2000).

For the association mapping, we transformed the phenotypes using a logarithmic transformation, which yields values that generally cause extreme late flowering plants to be less extreme. We then used AMM to map the phenotype, resulting in several interesting regions (Figure 6), including ones harboring known flowering genes such as *FRI* (Johanson *et al.*, 2000), *FLC*, *FLOWERING LOCUS T* (*FT*) (Shindo *et al.*, 2005; Huang *et al.*, 2005), and *DELAY OF GERMINATION 1* (*DOG1*) (Alonso-Blanco *et al.*, 2003; Bentsink *et al.*, 2006) (although *DOG1* is not a traditional candidate gene for flowering time it has repeatedly been suggested to affect flowering time in recent GWAS studies (Atwell *et al.*, 2010; Brachi *et al.*, 2010; Li *et al.*, 2010)).

Zooming in on the *FRI* gene (chromosome 4, position 269025-271503), we conditioned on the most significant SNP within the gene (position 269260, -log(p) = 4.8). This results in a very different association landscape around the *FRI* region, causing other SNPs near *FRI* to become significant at the 5% FDR threshold, where the SNP with the genome-wide smallest p-value (neg. log p-value 7.7) was at position 264496, less than 5kb from the transcription start site of *FRI* (Figure 7). After adding this SNP to the model as a cofactor, both SNPs become significant at the 5% FDR threshold, and appear to explain most of the signal in the region. These results are consistent with what is known about the role of *FRI* in flowering time (Shindo *et al.*, 2005; Aranzana *et al.*, 2005), i.e., there are at least two segregating variants within and near *FRI* that affect flowering time. Given that the indels are in negative linkage disequilibrium, it is not surprising that the signal becomes more pronounced after conditioning on one SNP within *FRI* (Atwell *et al.*, 2010; Platt *et al.*, 2010). Although the two known causal variants are not included in the dataset, because they are indels not SNPs, our analysis is still consistent with the known allelic heterogeneity.

We also included the most significant SNPs near known candidate genes, *FLC*, *FT*, and *DOG1*, in the model as cofactors. By doing so, many of the remaining peaks observed in the original scan dissipated (Supplementary Figure 5), leading us to believe that they were synthetic associations (Dickson *et al.*, 2010, Platt *et al.*, 2010). However, two peaks on chromosome 5 still remain. The first of these is located in the pericentromeric region (12.51-12.56 Mb), and only contains a handful of genes, none of which are obvious candidates (Supplementary Table 1). The second region (25.34-25.40 Mb) spans roughly 60 kb and includes SNPs with genome-wide significant p-values. This region does not contain any obvious candidate genes (Supplementary Table 2), however, it overlaps with a quantitative trait locus region recently observed for flowering time (Salomé *et al.*, 2011).

Finally, the mixed model estimated the narrow-sense heritability of flowering time to be 100%, which may seem extreme, but is actually not far from the more robust broad-sense estimate of 92%. The five cofactors included in the model explained 43% of the phenotypic variance. The estimated fraction of remaining genetic variance was 57%, and the estimated fraction of remaining error variance was 0% (Figure 8). This indicates that there are still unexplained genetic effects in the genome, with the two remaining peaks on chromosome 5 as prime candidate regions.

# Discussion

This paper is part of our overall effort to enable the *Arabidopsis* community to capitalize on the unique resources of thousands of densely genotyped lines. Over 1,300 lines have been genotyped using a 250k SNP chip (Horton *et al.*, 2012), and a thousand more will be sequenced by the end of this year (Gan *et al.*, 2011; Cao *et al.*, 2011; www.1001genomes.org). It is our hope that these lines will be routinely phenotyped to reveal functionally important polymorphisms via GWAS. One obstacle to this becoming a reality is the difficulty of analyzing the data: overcoming this difficulty is the direct objective of the work presented here.

Our goal was to provide an easy-to-use tool for GWAS that enables users to focus on biology instead of spending time programming or converting file formats. All that is required is a simple import of the phenotypic data, which can easily be managed in a spreadsheet. GWAPP provides several interactive features, including the possibility of analyzing different subsets of the sample as well as some basic transformations of the raw phenotypic data. With interactive Manhattan and genome annotation plots, it is possible to browse through the results, zoom in on association peaks, and quickly gain an overview of what genes may harbor causal variants. Patterns of LD can analysed, both local LD patterns as well as genome-wide LD patterns, that are calculated on-the-fly. To further aid interpretation, several statistics, including recombination rate and selection statistics, can be plotted along the chromosome. Conditional analysis using SNPs as cofactors makes it possible to investigate genetic heterogeneity, and estimates of variance components provide insight into the genetic architecture of the traits (Yang *et al.*, 2010). Furthermore, GWAPP can do all this in minutes. Using similarly sized datasets for benchmarking, as used for the benchmarks in *Matapax* (Childs *et al.*, 2012), we observed up to 50-fold increase in speed for a mixed model analysis of a single trait (see materials and methods).

To demonstrate how one might use GWAPP, we reanalyzed a previously published flowering time phenotype dataset (Li *et al.*, 2010). By leveraging *a priori* biological knowledge we identified two independent loci near *FRI,* which when included in the mixed model explain a quarter of the total phenotypic variance. After including associated SNPs near four genes known to be involved in flowering time, there were still loci of potential interest. Interestingly, one of these is in a region that was recently shown to be associated with flowering time in a linkage mapping study (Salomé *et al.*, 2011).

The web application presented here, GWAPP, is a work in progress. It can be extended in several ways and we are actively working on this. Most obviously, we will continuously increase the SNP data set by including overlapping SNP data from newly sequenced accessions (Cao *et al.*, 2011; Gao *et al.*, 2011). We will of course make it possible to utilized 'full' sequence data from the 1,001 genomes project, but this will require optimizations in order to run in real time. Another major improvement will be the ability to look for pleiotropy by looking for associations across all published phenotypic data. With the cooperation of the *Arabidopsis* community, it should be possible to establish a database that aims to functionally annotate every segregating polymorphism in the genome.

More trivially, the interface, tools, and methods can easily be changed, updated and expanded based on user input. Finally, although GWAPP is currently dedicated to GWAS in *A. thaliana,* some parts of the application, including the interactive plots and the underlying data-structures and mapping algorithms, can readily applied to data from other organisms, including humans. By structuring the application in modules, certain parts, e.g., interactive visualization components or the association mapping algorithms, can be easily reused for other projects. Importantly, all source code for our application is freely available.

## Materials and Methods

### Genotype data

The genotype data used was obtained by combining data from two different sources, namely 1386 *A. thaliana* accessions that were genotyped for 214K SNPs (Horton *et al.*, 2012), and 80 *A. thaliana* accessions that were sequenced using next generation sequencing (Cao *et al.*, 2011). One accession (*Fei-0*) was characterized in both analyses (n=1386) and we used the SNP calls from Horton *et al.* to correct the discordant SNP assignments (discordant rate was 2.5%). For the sequence data we extracted the base calls corresponding to the 214K SNP positions from the combined matrix, and imputed the missing alleles with *BEAGLE* version 3.3.1 (Browning and Browning, 2011). We used 30 iterations for the imputation, with the full merged dataset as phased input. All tri-allelic SNPs were discarded for simplicity, leaving 206,087 SNPs in the final dataset. The coordinates shown in the browser are TAIR 10 coordinates. GWAPP does not provide any easy way to upload custom genotype data yet. However users can download the virtual machine (VM) image of the application (see section VM image) and replace or extend the provided genotype data with custom ones.

## Web application

In order to minimize installation time and allow for widespread access we implemented GWAPP as a web application. The only client-side requirement is a browser that supports HTML5. The application consists of a back-end, front-end, and data exchange protocol (Supplementary Figure 6). The server front-end is the part of GWAPP that the user interacts with. The user interface and all visualization tools used for the analysis are a part of the front-end. The front-end is primarily implemented using modern web technologies (HTML5 and Javascript), The back-end implements all association mapping methods, various statistics, and performs all handling of data, such as parsing, coordination, and filtering of phenotypes and genotypes. The backend is written almost entirely in Python and is all server-side. Finally, the data exchange protocol is communicates between the front-end and the back-end. The implementation details for the server, i.e. front-end, back-end and the data exchange protocol, are described in (Supplemental Methods).

### VM Image

Since genotype data is typically large in size, GWAPP does not support uploading custom genotype data. Instead, we provide a shrink-wrapped package that has a version of GWAPP and all dependencies preinstalled and preconfigured as a VM image. The package also includes all non-standard packages necessary for installation and deployment of GWAPP on either on-premise/private cloud or public cloud services. The VM Image can be downloaded here:
https://cynin.gmi.oeaw.ac.at/home/resources/gwapp/gwapp
Further information for installing GWAPP locally is provided there, including information on how to use a different genotype dataset than the Horton *et al.* dataset (Horton *et al.*, 2012).

## Mapping methods

Three different mapping methods were implemented for GWAPP. A standard linear regression (LM), an approximate mixed model (AMM) (Kang *et al.*, 2010; Zhang *et al.*, 2010), and a Wilcoxon rank-sum test (Wilcoxon, 1945). AMM differs slightly from EMMAX (Kang *et al.*, 2010) and P3D (Zhang *et al.*, 2010), in that it re-estimates the p-values for the k=100 most significant SNPs using exact inference (Kang *et al.*, 2008; Lippert *et al.*, 2011). The exact inference re-estimates the variance components with the SNP in the model as a cofactor, and then uses these updated variance components to re-estimate the p-value of that SNP. AMM has running time complexity of $O(n^2m+n^3k)$, where *n* denotes the number of individuals, and *m* the number of SNPs. If we choose $k \leq m/n$, the running time becomes $O(n^2m)$, i.e. the same as EMMAX (Kang et al., 2012), FaST-LMM (Lippert *et al.*, 2011), and GEMMA (Zhou and Stephens, 2012). Furthermore, LM and AMM were implemented to allow for inclusion of SNPs into the model (Segura *et al.*, 2012), using the Gram–Schmidt process to ensure efficiency regardless of the number of cofactors included. See online methods in Segura *et al.* (Segura *et al.*, 2012) for further details. The three mapping methods were implemented in Python by extending *mixmogam* (https://github.com/bvilhjal/mixmogam) (Segura *et al.*, 2012). We compiled *SciPy* (Jones *et al.*, 2001) with the *GotoBlas2* (Goto *et al.*, 2008) *Basic Linear Algebra Subroutines* (BLAS) implementation on the publicly available web server (https://gwas.gmi.oeaw.ac.at). For AMM, the genetic relatedness matrix used is the identity by state (IBS) genetic relatedness matrix, which for a pair of individuals is the fraction of shared alleles among segregating SNPs in the sample. This is calculated a priori for the full genotype dataset, and then adjusted for each specific subset of accessions by removing the contributions of SNPs, which are not segregating the subset (monomorphic SNPs).

### Runtime analysis

To benchmark the performance of GWAPP, six datasets were generated using random phenotype values (sampled from a uniform distribution), and using all 214K SNPs. The benchmark was conducted on the public web server, where GotoBlas2 was configured to use up to four cores (which is used by AMM and LM for linear algebra operations). The time was measured from pressing the analysis method button until all the p-values were displayed in the Manhattan plots. All three mapping methods finished within five minutes for all the datasets (Figure 9). AMM was considerably slower than the other two, but all methods finished the analysis within two minutes when using less than 500 individuals. This is about 50 times faster than *Matapax* (Childs *et al.*, 2012), which required more than one hour to run on 500 individuals using a single trait and a similar sized genotype dataset.

## Accession Numbers

The four candidate genes for flowering time have the following *Arabidopsis* Genome Initiative locus identifiers: *FRI* (*At4g00650*), *FLC* (*At5g10140*), *FT* (*At1g65480*), and *DOG1* (*At5g45830*). The genotype data can be found here: https://cynin.gmi.oeaw.ac.at/home/resources/atpolydb/250k-snp-data/call_method_75.tar.gz

## Supplemental Data

**Supplementary Figure 1.** Analysis levels of GWAPP.
**Supplementary Figure 2.** LD visualization – highlighting a SNP.
**Supplementary Figure 3.** LD visualization – highlighting a r² value
**Supplementary Figure 4.** Genome-wide LD visualization.
**Supplementary Figure 5.** AMM scan after conditioning on five SNPs.
**Supplementary Figure 6.** Overview of the web application structure.
**Supplementary Table 1.** Genes located in a region (12.51-12.56 Mb) on chromosome 5, which displayed association with flowering time.
**Supplementary Table 2.** Genes located in a 60kb region (25.34-25.39 Mb) on chromosome 5, which displayed association with flowering time.
**Supplementary Methods.** GWAPP implementation details.

## Acknowledgements

We would like to thank Arthur Korte, Envel Kerdaffrec, Fernando Rabanal, Wolfgang Busch, Danielle Filiault and others for testing the software and providing feedback. We would also like to thank Geoff Clarck for his useful comments. This work was supported by grants from the EU Framework Programme 7 ('TransPLANT', grant agreement number 283496) to M.N., as well as by the Austrian Academy of Sciences through the GMI.

## Author Contributions

Ü.S., B.J.V., V.S., and M.N. designed the study. Ü.S. and B.J.V. implemented and coded the web application. M.W.H. and Q.L. provided genome-wide statistics. D.M. and B.J.V. prepared the genotype data. P.F. and Ü.S. set up the server. Ü.S., B.J.V, and M.N. wrote the paper with input from all authors.

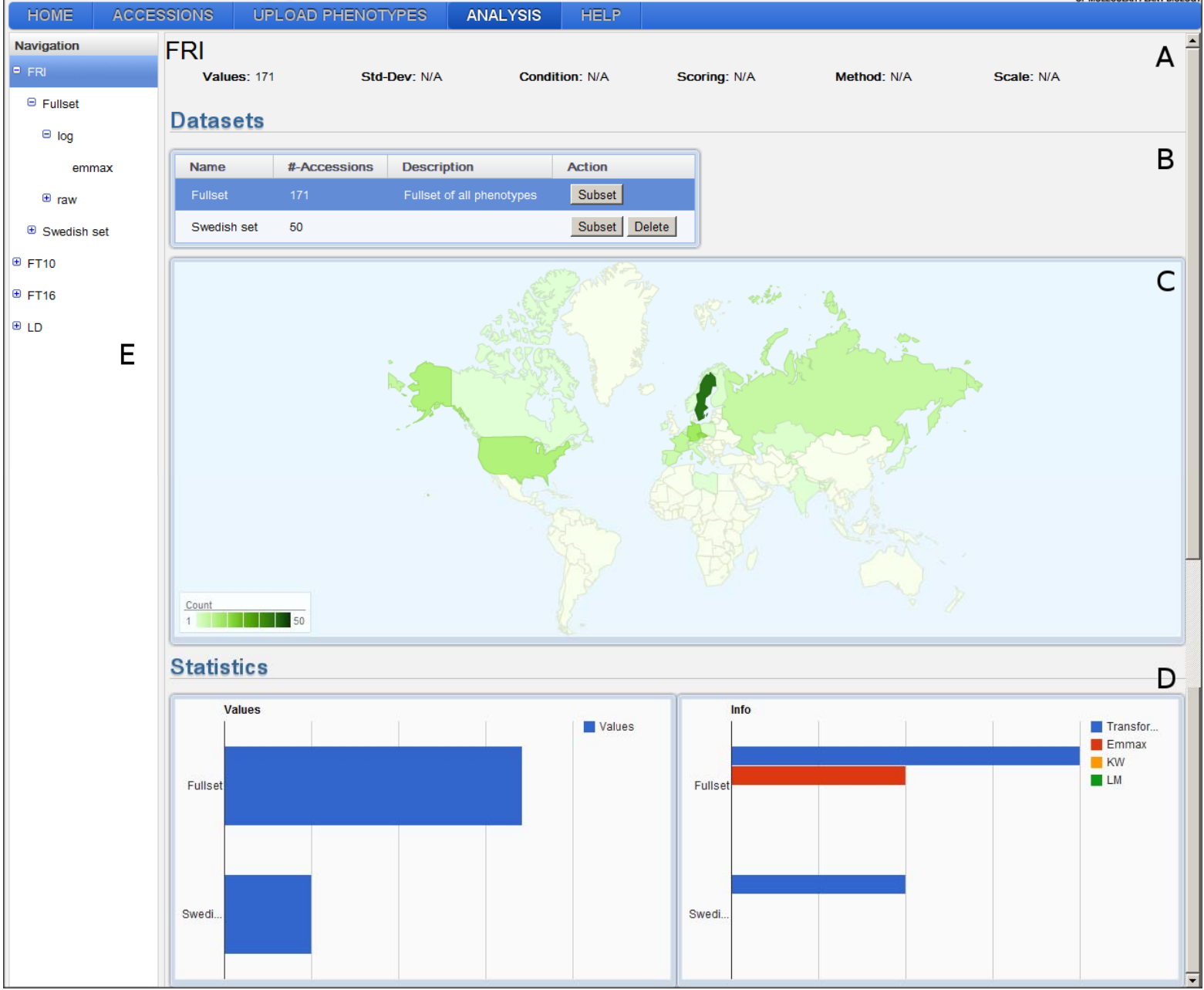

**Figure 1.** Phenotype view.
The phenotype view shows phenotype specific information in four panels. The top panel (**A**) displays phenotype name and number of values. In the panel below (**B**) a list of datasets is shown. Selecting a dataset from that list will update the geographic distribution map (**C**). Two bar charts in the bottom panel (**D**) show statistical information about the phenotype. The navigation tree on the left side (**E**) reflects the stored phenotype structure and is used to access different views

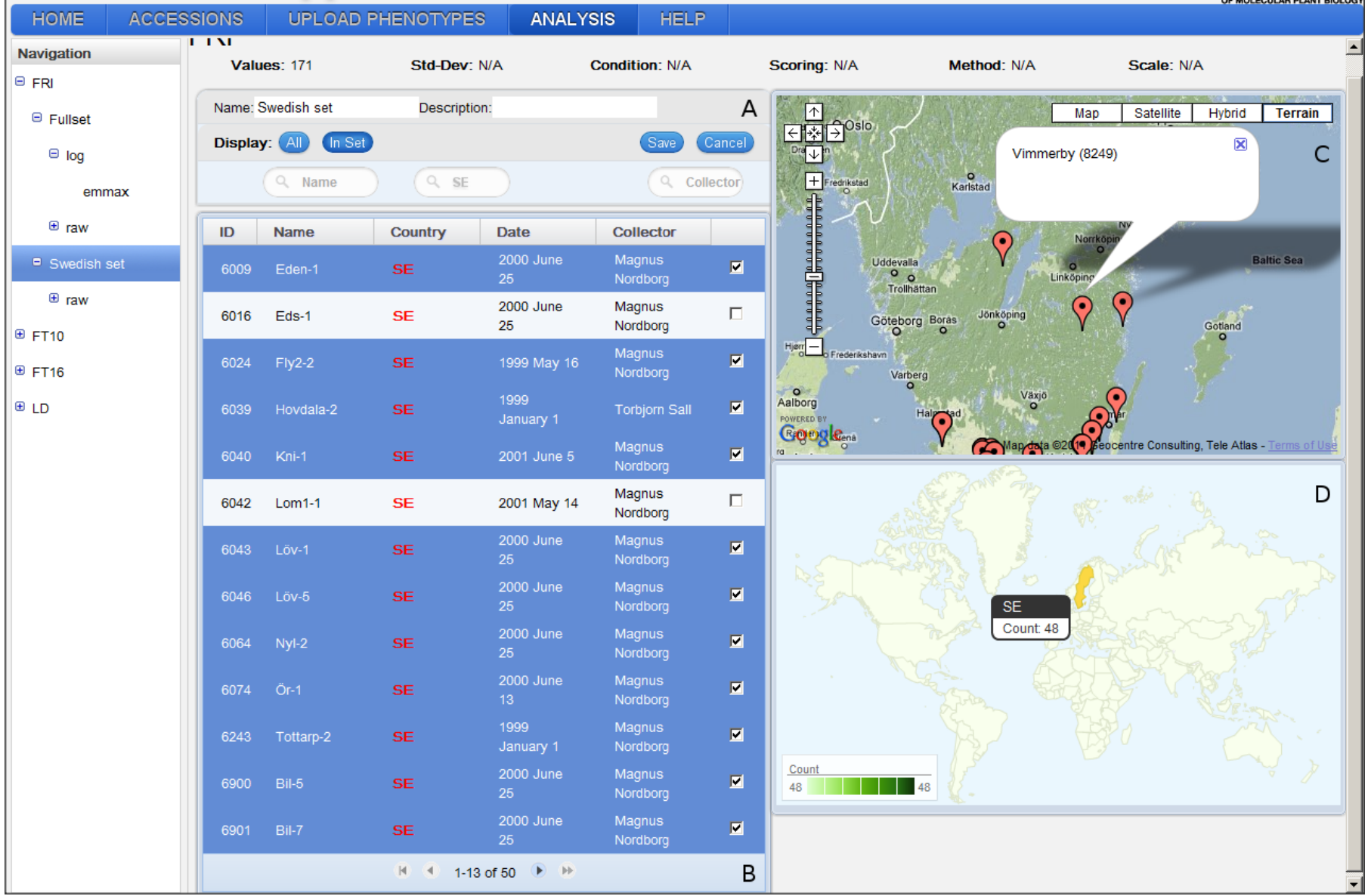

**Figure 2**. Dataset view.
(**A**) The 'filter box' provides filter functionality for the dataset list and allows the user to change the dataset name and description. (**B**) the dataset list shows information about the accessions in the dataset. In edit mode the user can use the checkbox to add and remove accessions from the dataset. (**C**) A Google map shows the locations of all accessions in the dataset. Clicking on one marker will show a popup with information about the name and id of the selected accession. The geographic distribution map (GeoMap) shows the geographic distribution of the accessions in the dataset (**D**). Moving the mouse over a country will show the number of accessions located in that region.

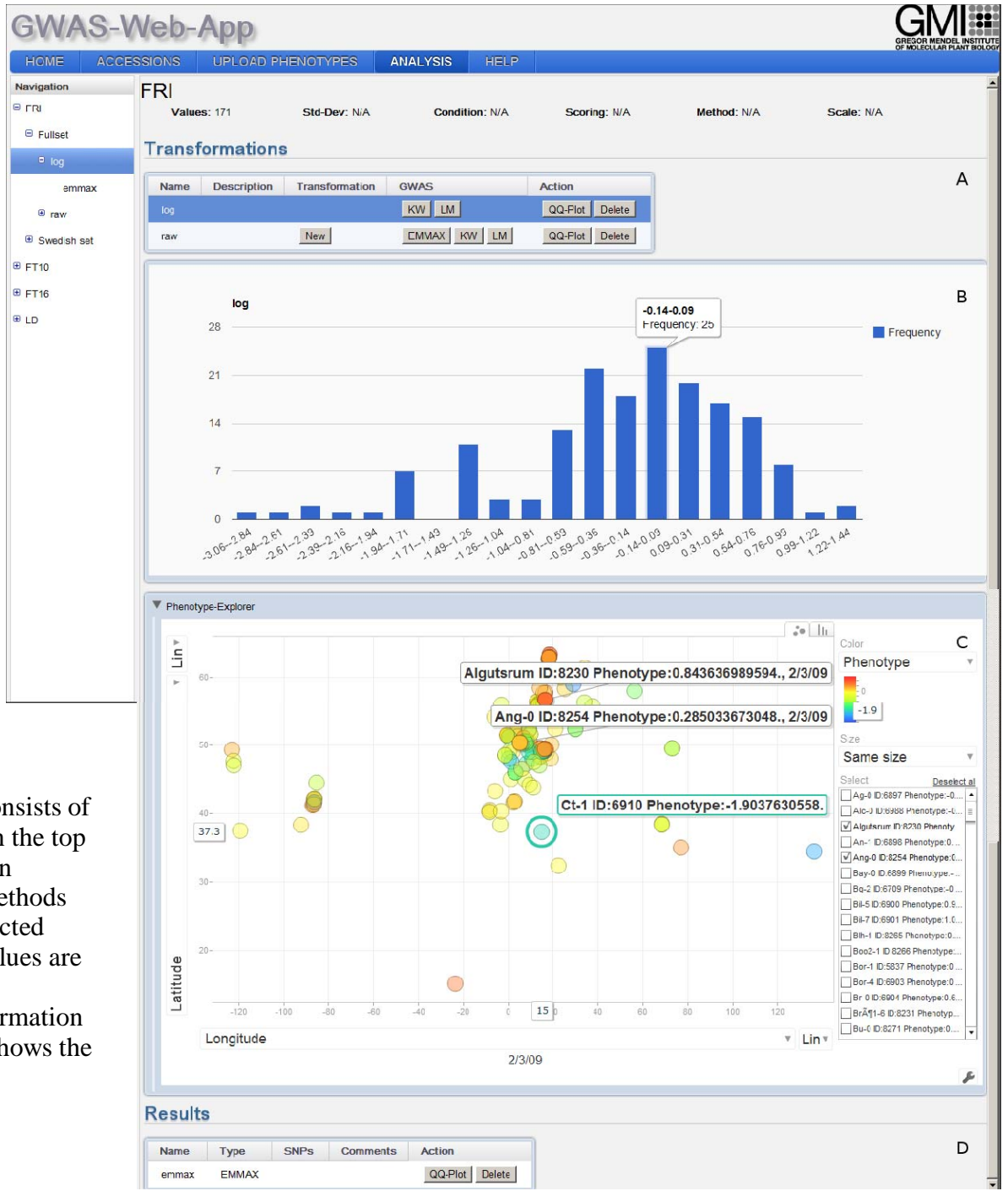

**Figure 3.** Transformation view. The transformation view consists of four panels. The list of stored transformations is displayed in the top panel (**A**). The use can create a new transformation, delete an existing one or run one of three available GWAS analysis methods on the transformed phenotype values. Dependent on the selected transformation a histogram of the transformed phenotype values are displayed below the transformation list (**B**). The Accession-Phenotype-Explorer (**C**) visualizes additional accession information through a bar-chart or a scatter-plot. The bottom panel (**D**) shows the stored GWAS-results for the specific transformation.

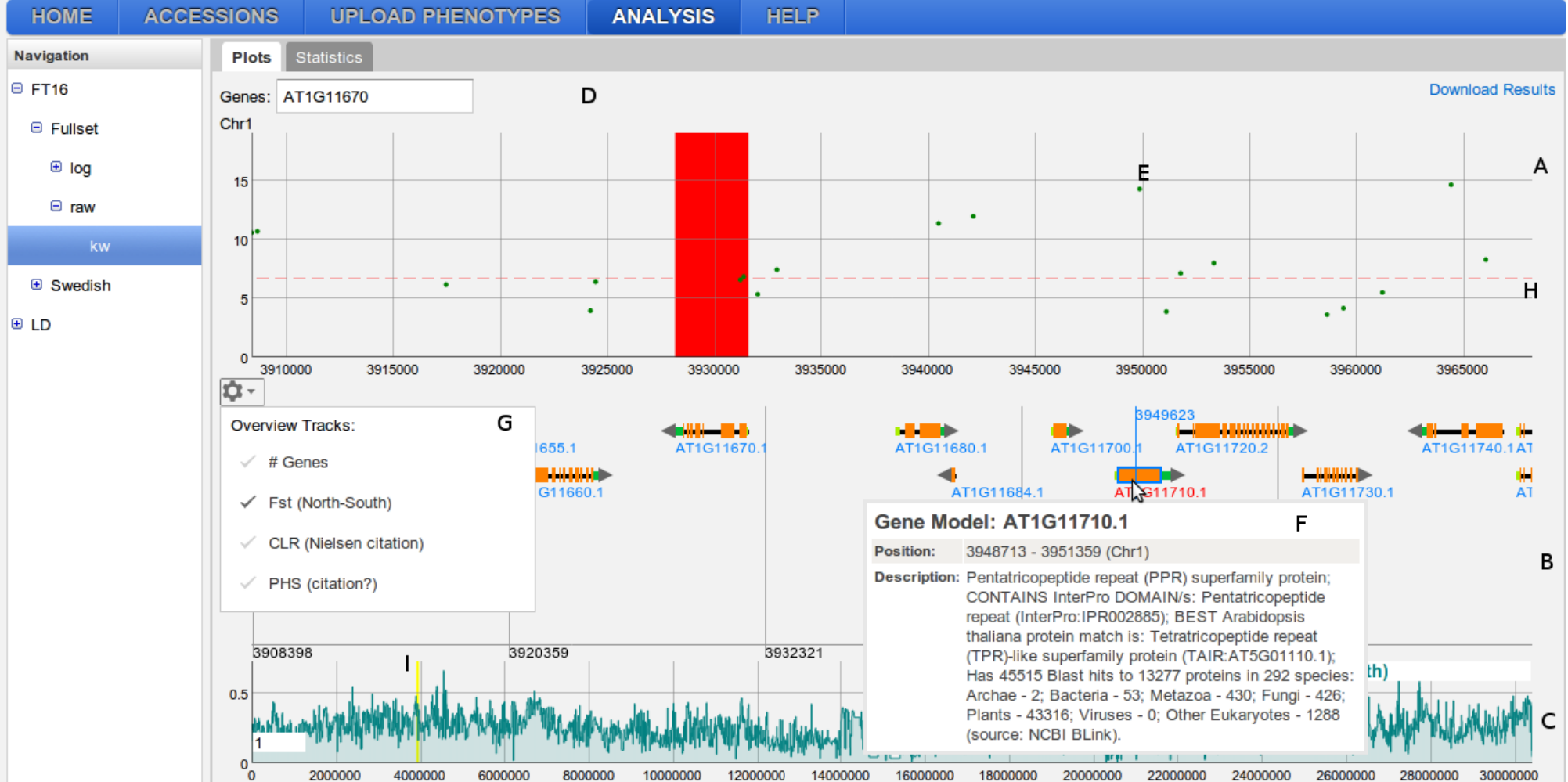

**Figure 4.** Result view.
The result view displays GWAS-plots for each of the five chromosomes. Each GWAS-plot itself consists of three panels. The top panel (**A**) contains a scatter-plot. The positions on the chromosome are on the x-axis and the score on the y-axis. The dots in the scatter-plot represent SNPs (**E**). A horizontal dashed line (**H**) shows the 5% FRD threshold. At the top of the GWAS-results view, a search-box for genes is displayed (**D**). These genes will be displayed as a colored band (red in the figure). The second panel (**B**) shows the gene annotation and is only shown for a specific zoom-range (< 1.5 Mpb). It will display genes, gene features and gene names. Moving the mouse over a gene will display additional information in a popup (**F**) and clicking on a gene will open the TAIR page for the specific gene. The bottom panel (**C**) displays various chromosome-wide statistics. The region shown in the scatter-plot and the gene annotation view is highlighted by a yellow band (**I**). The gear icon opens a popup (**G**) with the available statistics the user can choose from.

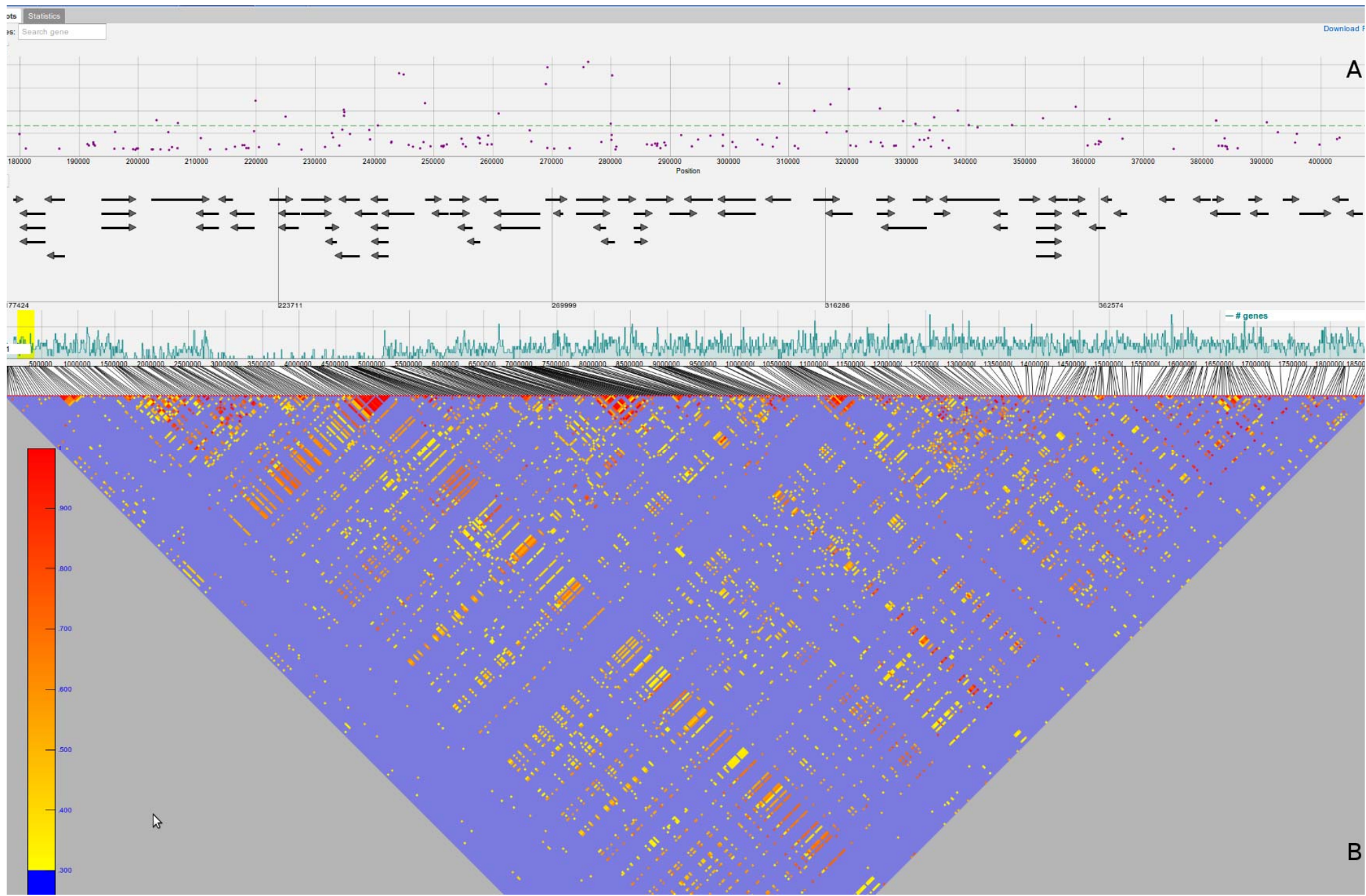

**Figure 5.** LD visualization
A screenshot showing the visualization of Linkage Disequilibrium for a specific region (500 SNPs). The triangle plot below (**B**) the gene annotation pannel shows the r² values of 500 SNPs. Only r² value above are color-coded ranging from yellow (low) to red (high).

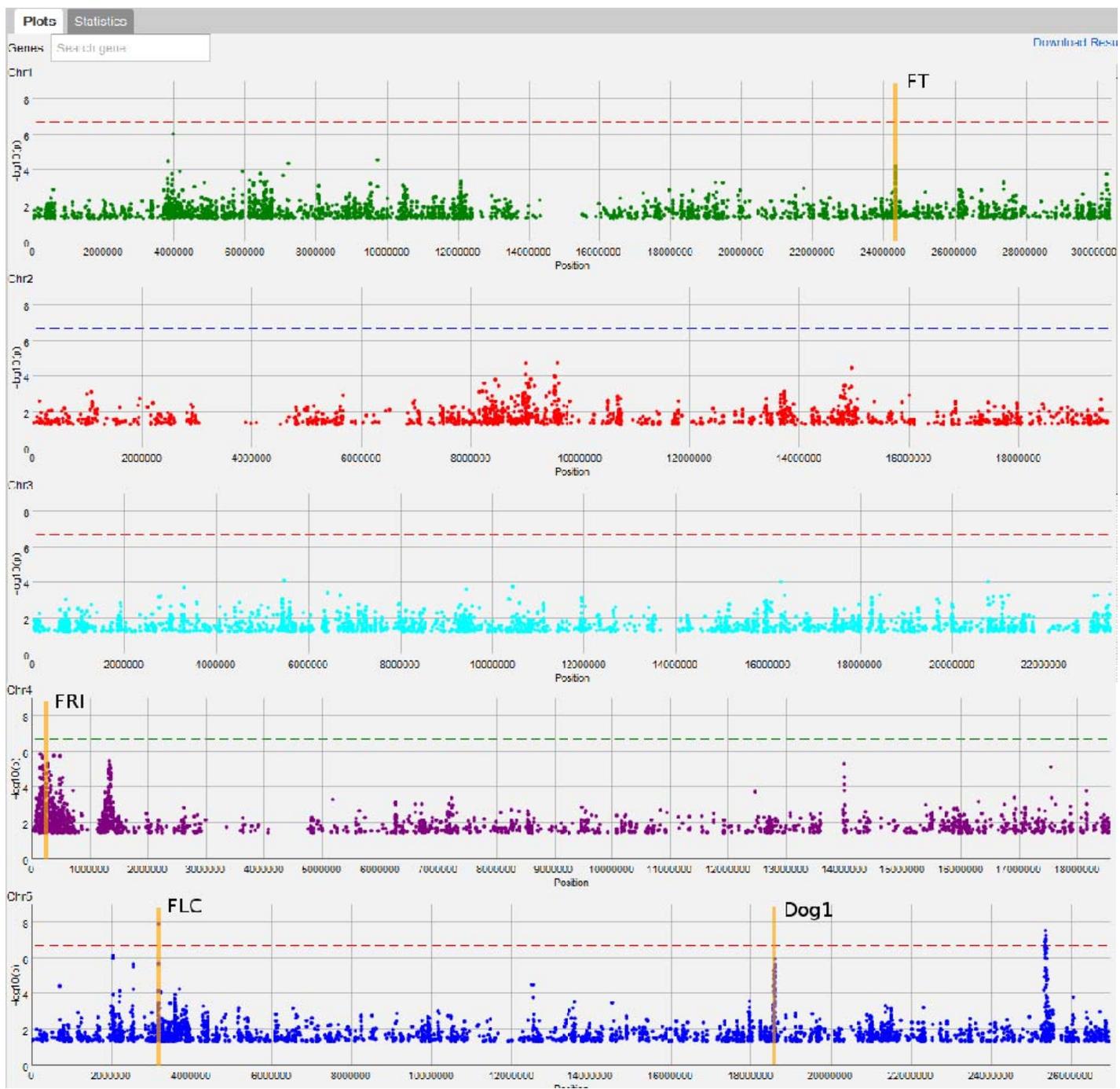

**Figure 6.** First AMM scan for flowering time.
A screenshot showing the first mixed-model scan for flowering time, highlighting the positions of four interesting candidate genes (*FT*, *FRI*, *FLC*, and *DOG1*) for which there seem to be associations..

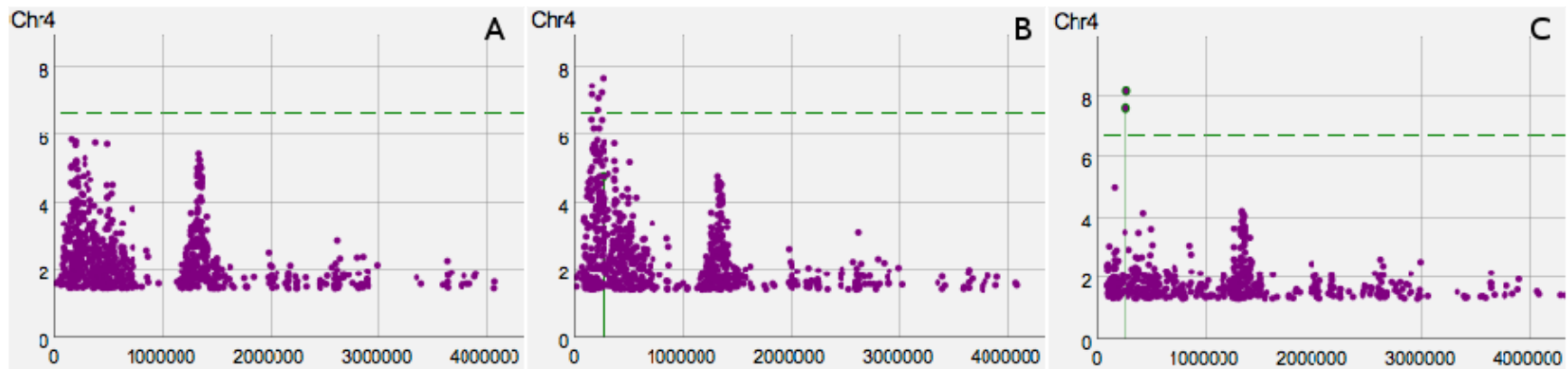

**Figure 7.** Conditional mixed-model scans for flowering time.
The first AMM scan (**A**) without any cofactors is shown on the left. The second AMM scan (**B**) in the middle, is the result from adding the SNP with the smallest p-value within the *FRI* gene into the model as a cofactor. Finally the third AMM scan (**C**) on the right, is the result from adding the top SNP from the middle figure, which is 5kb upstream of the *FRI* gene into the model as a cofactor. The negative log p-values are shown on the y-axis and the positions on the x-axis. The 5% FDR threshold is denoted by a horizontal dashed green line.

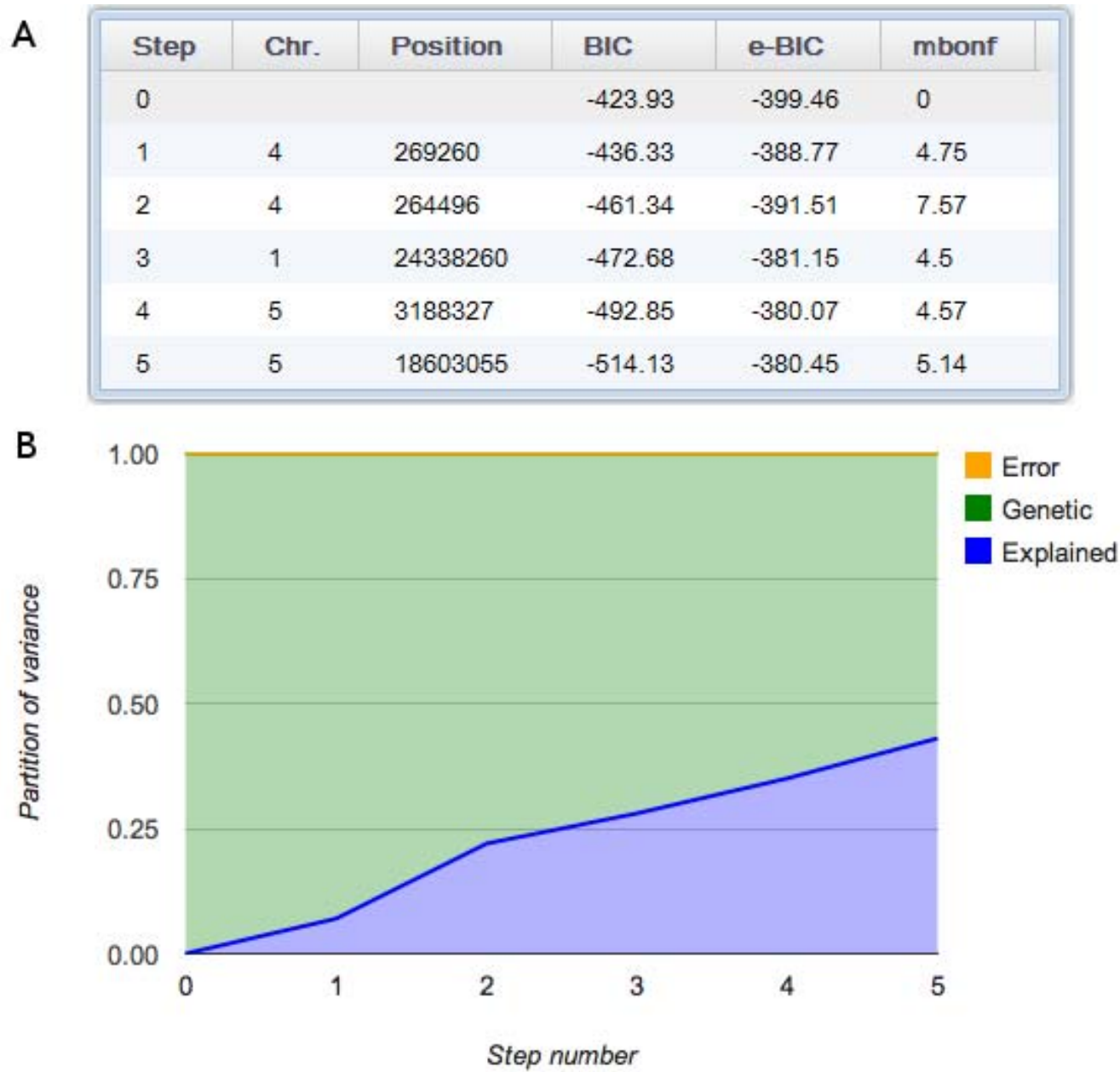

**Figure 8.** Partition of variance for the conditional mixed-model scans.
Two screenshots showing the five SNPs included in the model (**A**) and how the partition of phenotypic variance changes as the five cofactors (*FRI*, *FT*, *FLC*, and *DOG1*) are added to the mixed model (**B**).

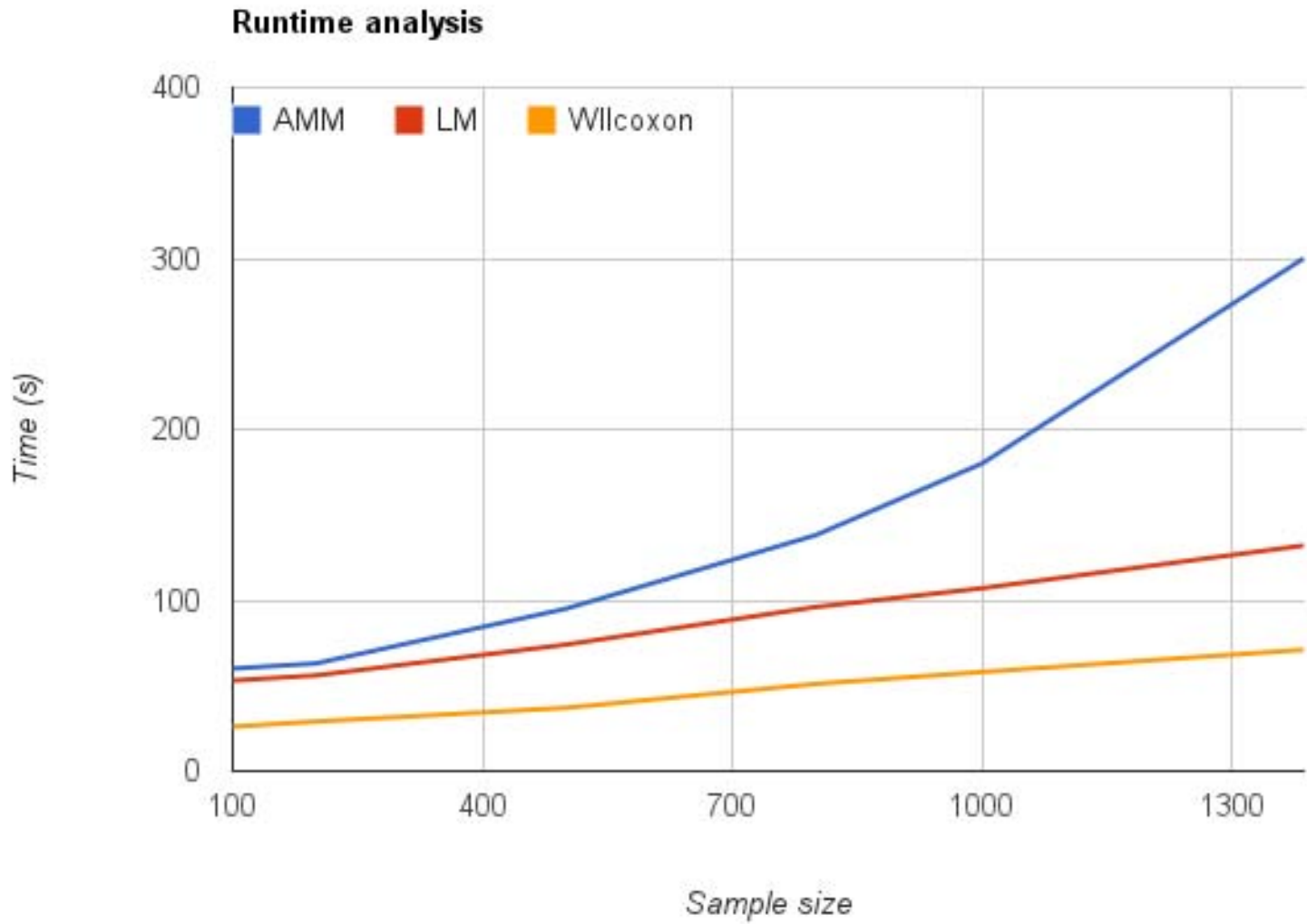

**Figure 9.** Runtime for different mapping methods.
The time, from starting the analysis until the p-values are visible in the Manhattan plot, is plotted against the number of individuals used for the GWAS. Lines for all three mapping methods are shown, AMM, LM, and the Wilcoxon rank-sum test.

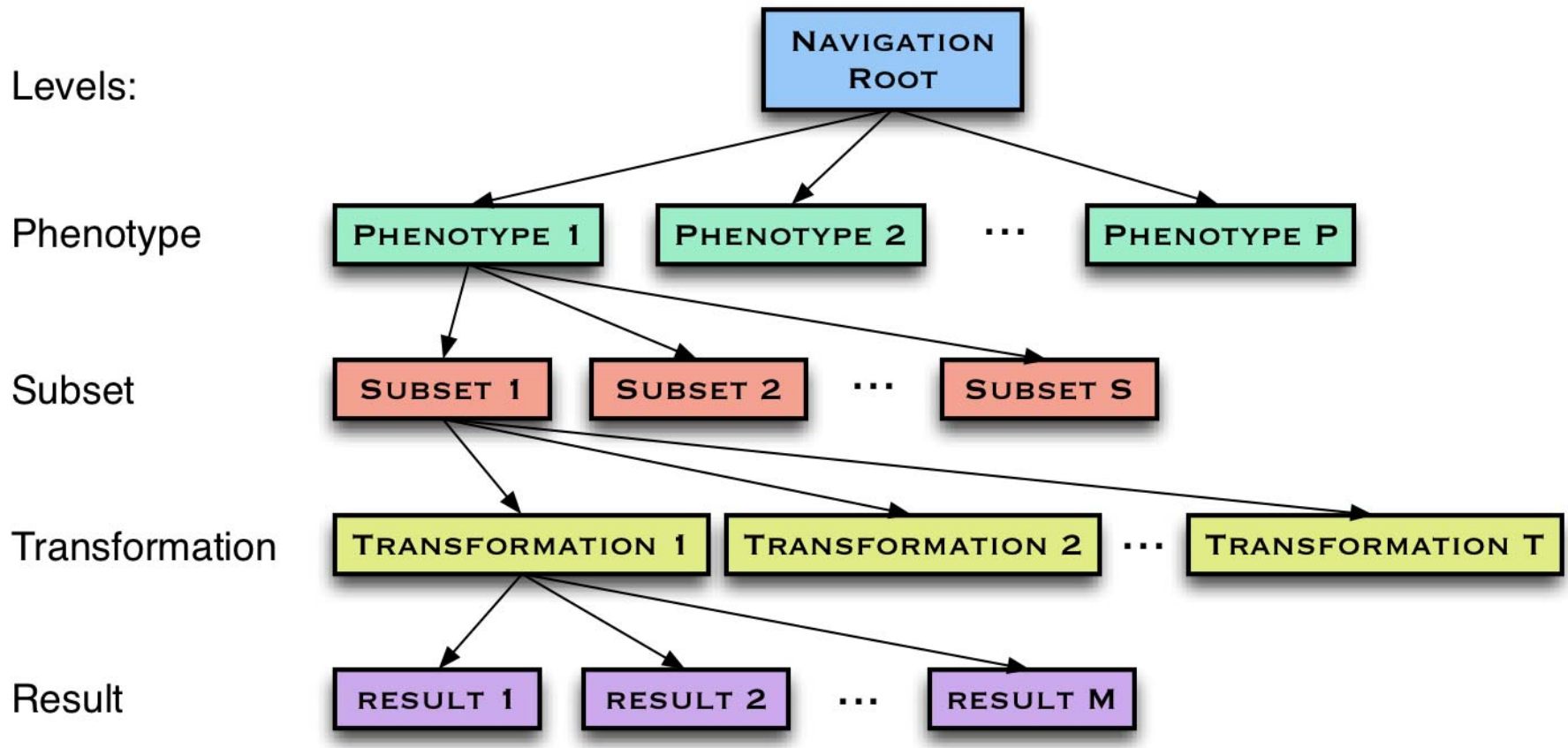

**Supplementary Figure 1.** Analysis levels of GWAPP. The four analysis levels, where the user starts by uploading phenotypes (**top**) and then continues towards obtaining mapping results (**bottom**).

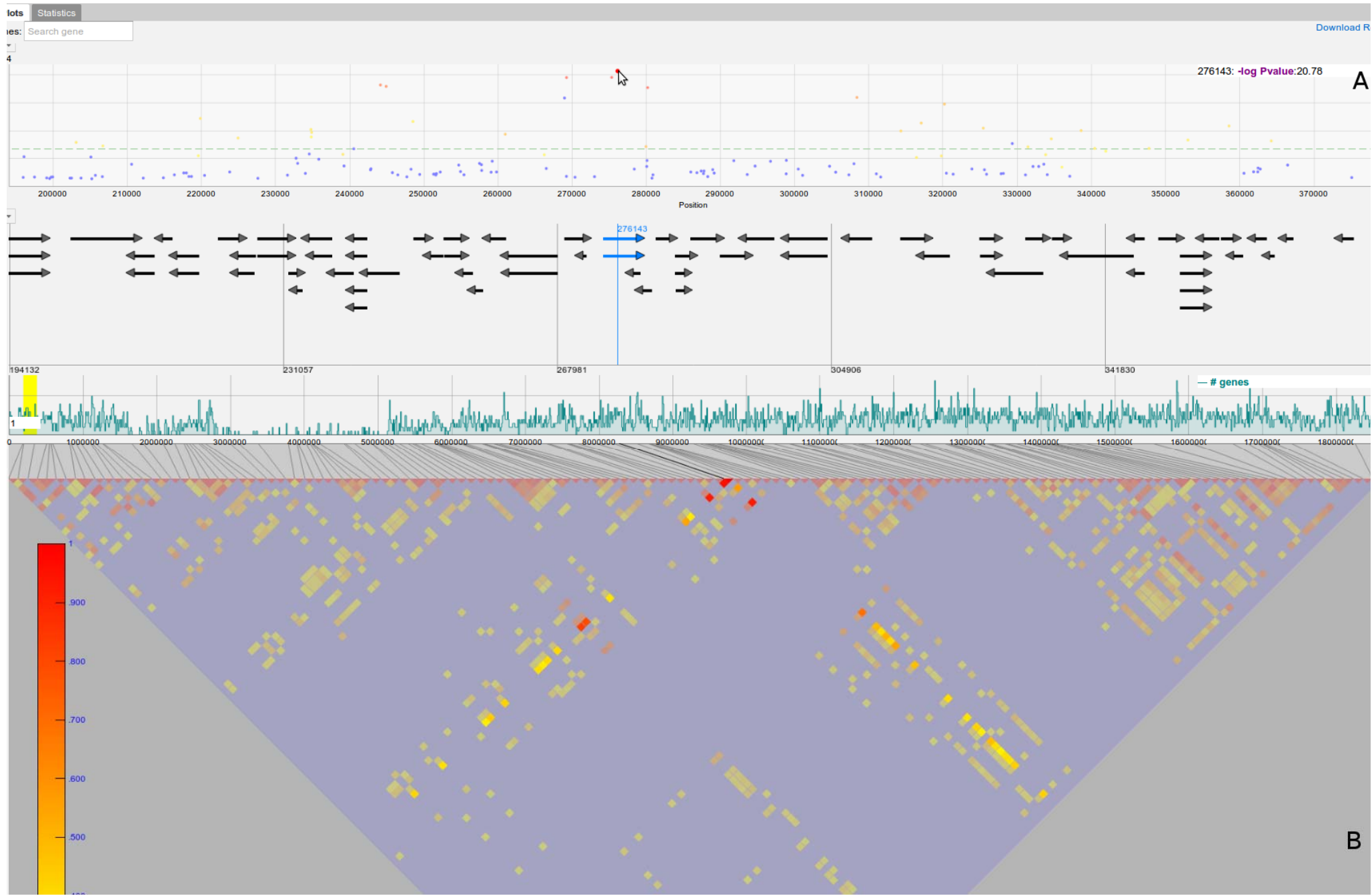

**Supplementary Figure 2.** LD visualization – highlighting a SNP
When a SNP is highlighted in the Manhattan plot (**A**), all neighbouring SNPs are color-coded acoording to their r² value and the corresponding r² values in the triangle plot (**B**) are highlighted with the corresponding color-coding.

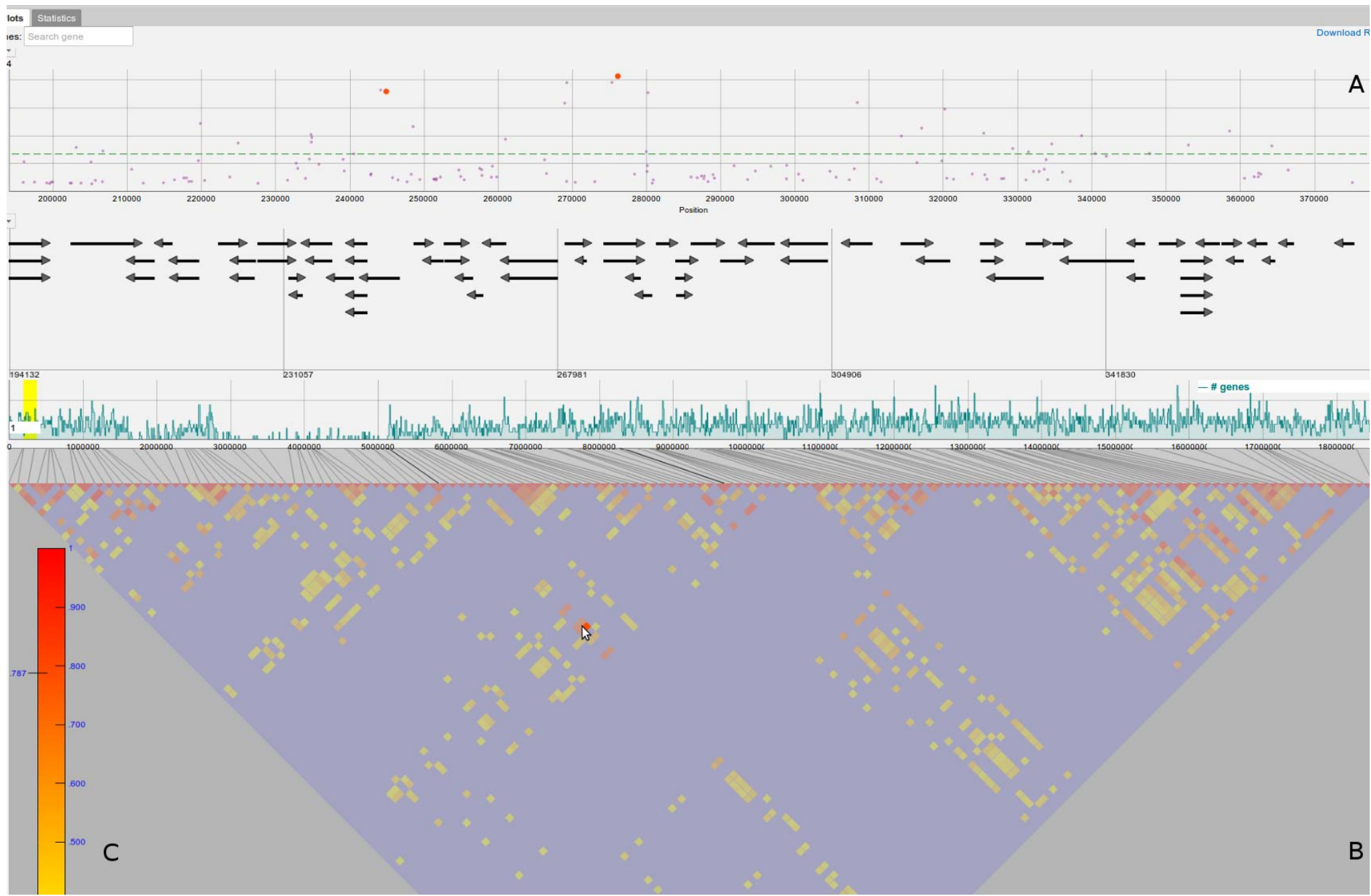

**Supplementary Figure 3.** LD visualization – highlighting a r² value
When a r² value of a SNP pair in the triangle plot (**B**) is highlighted, the two corresponding SNPs in the Manhattan plot (**A**) are color-coded according to the r² value on the scale (**C**).

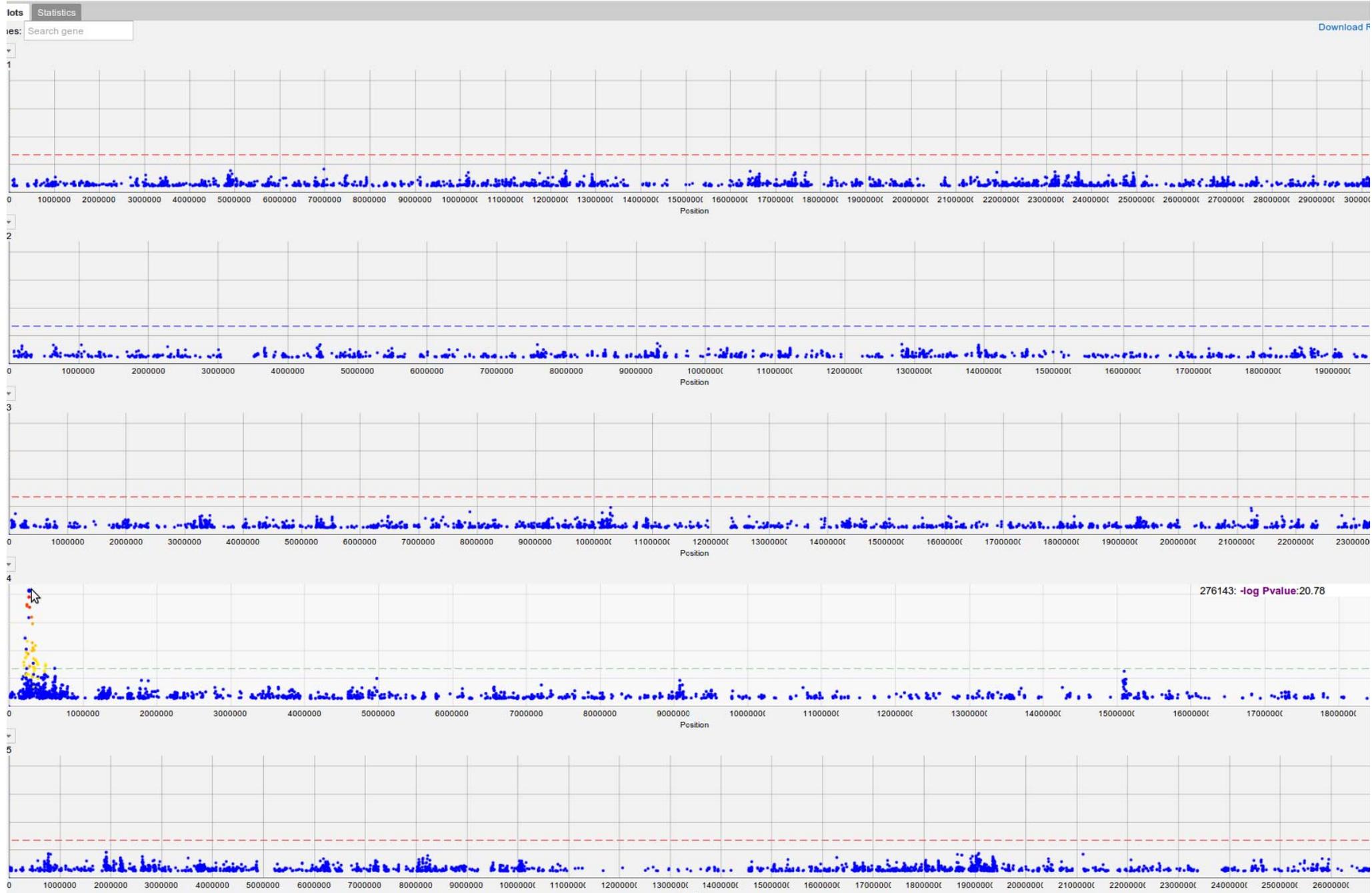

**Supplementary Figure 4.** Genome-wide LD visualization
Genome-wide LD for a selected SNP is displayed by color-coding all SNPs with r² values above 0.3.

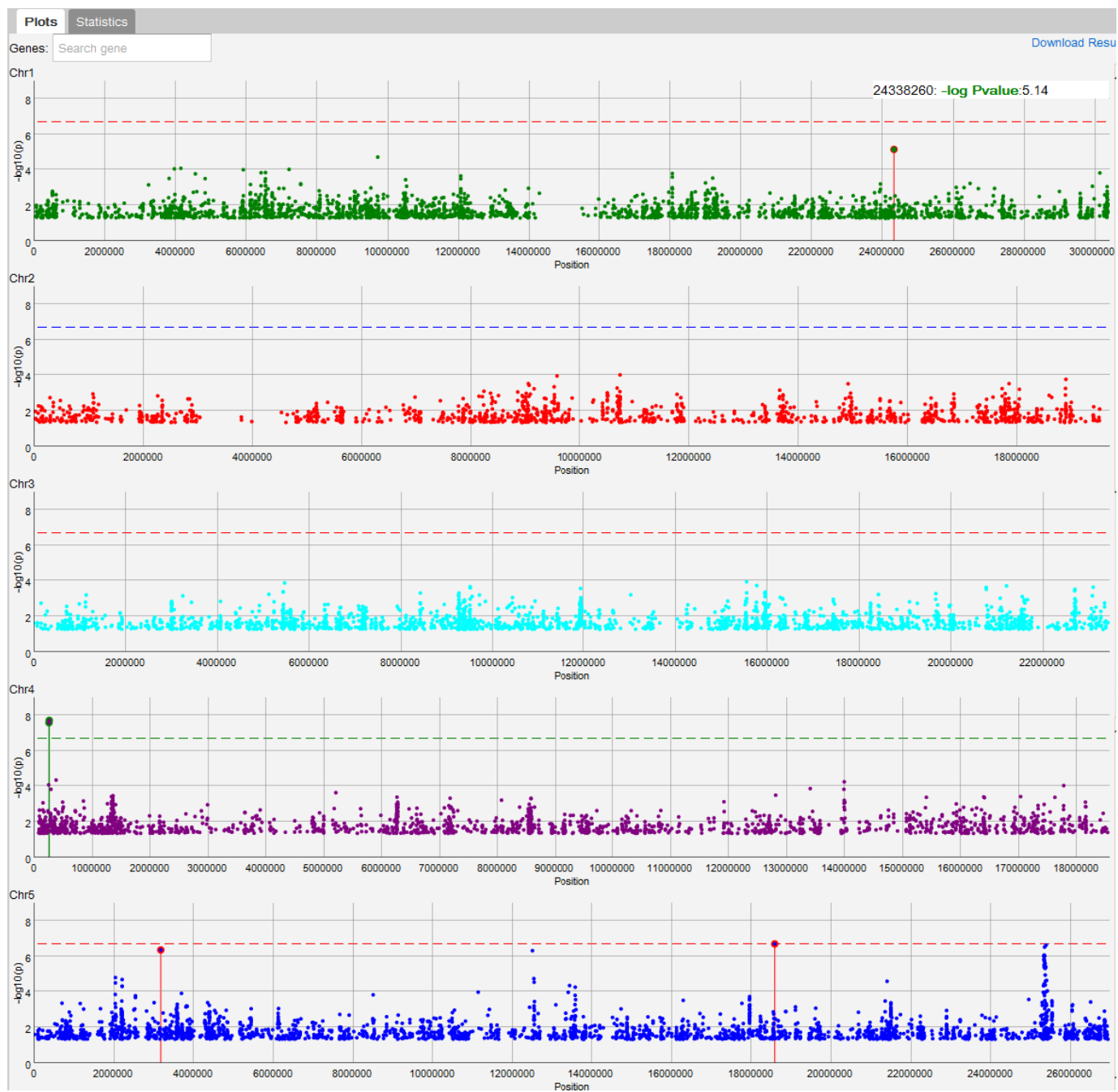

**Supplementary Figure 5.** Mixed-model scan after conditioning on five SNPs. A screenshot displaying the Manhattan plots after including the five SNPs near candidate genes (*FT*, *FRI*, *FLC*, and *DOG1*) as cofactors in the mixed model.

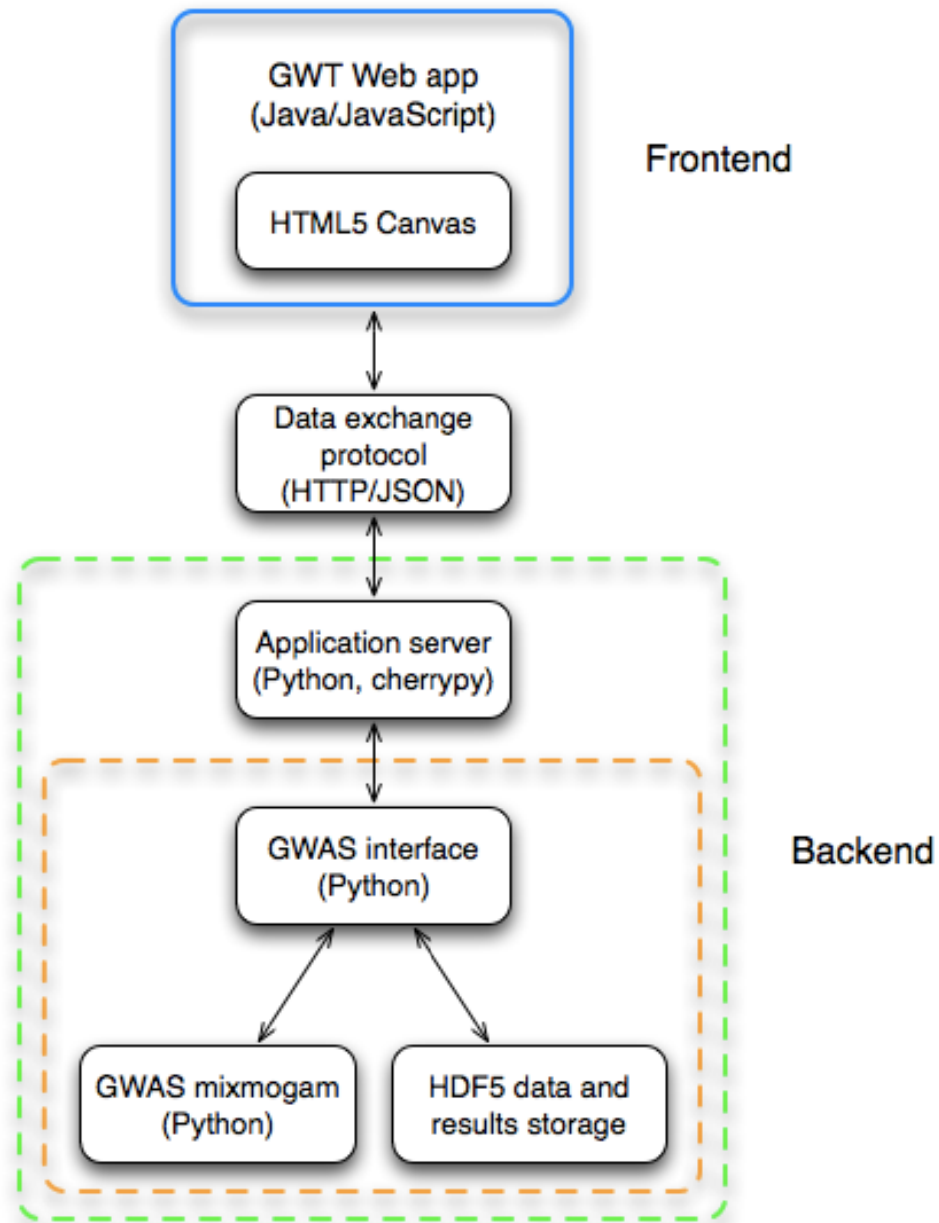

**Supplementary Figure 6.** Overview of the web application structure. GWAPP was designed in a modular way, consisting of a back-end and a front-end.

# Supplementary tables

| TAIR id | Start | End | Strand | Functional annotation |
|---|---|---|---|---|
| AT5G33280 | 12549280 | 12552507 | + | Voltage-gated chloride channel family protein; |
| AT5G33290 | 12558256 | 12562020 | + | Acts as a xylogalacturonan xylosyltransferase within the XGA biosynthesis pathway. |
| AT5G33300 | 12562430 | 12565757 | - | chromosome-associated kinesin-related; |

**Table 1.** Genes located in a region (12.51-12.56 Mb) on chromosome 5, which displayed association with flowering time.

| TAIR id | Start | End | Strand | Functional annotation |
|---|---|---|---|---|
| AT5G63190 | 25345542 | 25348796 | + | MA3 domain-containing protein |
| AT5G63200 | 25349011 | 25353039 | + | tetratricopeptide repeat (TPR)-containing protein |
| AT5G63220 | 25353034 | 25355500 | - | unknown protein; |
| AT5G63225 | 25356133 | 25356740 | - | glycosyl hydrolase family protein 17 |
| AT5G63230 | 25357360 | 25357822 | - | glycosyl hydrolase family protein 17 |
| AT5G63240 | 25358968 | 25359565 | - | Carbohydrate-binding X8 domain superfamily protein |
| AT5G63260 | 25361747 | 25364769 | + | Zinc finger C-x8-C-x5-C-x3-H type family protein |
| AT5G63270 | 25365498 | 25365949 | - | RPM1-interacting protein 4 (RIN4) family protein |
| AT5G63280 | 25367031 | 25369297 | + | C2H2-like zinc finger protein |
| AT5G63290 | 25369245 | 25370921 | - | CPO3 (At5g63290) has not been characterized per se, but is a homolog of the CPDH/HEMN |
| AT5G63300 | 25371122 | 25371775 | + | Ribosomal protein S21 family protein |
| AT5G63310 | 25371904 | 25373861 | - | Maintains intracellular dNTP levels except ATP |
| AT5G63320 | 25374413 | 25379246 | - | Encodes NPX1 (Nuclear Protein X1), a nuclear factor regulating abscisic acid responses. |
| AT5G63340 | 25378894 | 25379674 | + | unknown protein; |

| AT5G63350 | 25380000 | 25381183 | - | unknown protein; |
| AT5G63370 | 25383906 | 25387224 | - | Protein kinase superfamily protein; |
| AT5G63380 | 25387411 | 25390063 | - | Encodes a peroxisomal protein involved in the activation of fatty acids through esterification with CoA |
| AT5G63390 | 25390512 | 25392591 | - | O-fucosyltransferase family protein |

**Table 2.** Genes located in a 60kb region (25.34-25.39 Mb) on chromosome 5, which displayed association with flowering time.

# Supplementary Methods:

# GWAPP implementation details.

**Front-end**

The front-end was implemented using the *Google Web Toolkit* (GWT). GWT (http://code.google.com/webtoolkit/) is a development toolkit for building complex browser-based applications in Java. GWT translates the Java code to highly optimized Javascript code, which is run in the browser, and takes care of cross-browser incompatibilities.

Visualizations and interactive plots are key features of GWAPP, where interactive plots are only perceived interactive if they react promptly to user interactions. Recent developments in web- and browser-technologies enable fast web-based interactive plots and visualizations using two main techniques: (1) scalable vector graphics (SVG) and (2) HTML5 canvas. For the Manhattan plots we chose to use HTML5 as it allows for nearly native like performance in data intensive visualizations. GWAPP implements various different plots and charts that are displayed in the front-end, where the most important ones are arguably the Manhattan plots. For those we used Dan Vanderkam's open source javascript visualization library Dygraphs (http://dygraphs.com/). Dygraphs also supports out-of-the-box vertical and horizontal zooming, an important feature when analyzing GWAS results. Only the smallest %2 (~4,280) of all p-values are shown in the Manhattan plot to ensure rapid rendering. Furthermore, LM and AMM results are filtered for SNPs with minor allele count below 15, as rare SNPs can cause false positives for parametric tests. The full results can be downloaded by clicking on 'download result' in the upper right corner of the results view.

The gene annotation browser/viewer was implemented with the Processing.js library (http://processingjs.org), which relies on code from the Processing visualization programming language (http://processing.org) and renders it using web-technologies, specifically HTML5 Canvas. The data structure for the gene annotation browser is generated from TAIR10 using the data conversion scripts from *JBrowse* (Skinner *et al.*, 2009). Like *JBrowse*, we also make use of nested containment lists (NCList) (Alekseyenko *et al.*, 2007) for fast gene retrieval and interval queries, allowing for nearly instantaneous visualization of genes. Most charts and visualizations were created using the *Google Chart Tools* (http://code.google.com/apis/chart/) that uses SVG for rendering. As with HTML5 canvas, SVG is also supported by most modern browsers.

Finally, the front-end uses a caching mechanism to avoid unnecessary and redundant calls to the back-end, thus saving bandwidth and improving overall performance. For a list of all used components refer to the source code page of GWAPP http://code.google.com/p/gwas-web-app/.
levels (Supplementary Figure 1), the HDF5 file contains the four levels, where it stores different phenotypes, subsets, transformations and results.

For the Box-Cox transformation of the phenotype values, an optimal lambda value is chosen with respect to normality using a Shapiro-Wilks test. The lambda values considered range between -2 and 2 in increments of 0.1. As these transformations assume that the phenotype values are non-negative, the phenotypes are shifted prior to the transformation by subtracting the minimum phenotype value and adding 1/10 of the phenotype standard deviation.

When a GWAS is performed, the input data, genotype and phenotype, are first loaded and synchronized. The actual GWAS is conducted only after the input data has been synchronized and filtered appropriately. Phenotype replicates are averaged (according to our experience, allowing for replicates in the mixed model rarely improves power for detecting associations). After mapping, p-value quantiles are calculated and stored together with the results, which are ordered by score. By storing the p-values in order, the application can quickly load the top 2% of the p-values for the Manhattan plots.

**Back-end**

The back-end consists of three tiers: (1) the application server, (2) the GWAS interfaces, and (3) the *Hierarchical Data Format* version 5 (HDF5) (The HDF Group, 2000) data and result storage. We implemented the back-end entirely in Python. *CherryPy* (http://cherrypy.org), a pythonic object-oriented HTTP framework, was used as an application server. On the back-end side, the application server communicates with the GWAS interface which itself can either directly access the HDF5 data storage to retrieve data and results or can access the GWAS *mixmogam* (Segura *et al.*, 2012) package for processing data and performing the analysis.

PyTables (Alted *et al.*, 2002) was used for accessing the HDF5 data storage from Python. GWAPP solely relies on the HDF5 file for data storage and not any external database, thus ensuring portability. Each user has its own HDF5 file, which contains all phenotype data and the GWAS-results and is tied to a client-side cookie. The HDF5 file can be downloaded and viewed by any standard HDF5 viewer. Analogous the analysis

**Data exchange protocol**

As our web-application is based on a service oriented front-end architecture (SOFEA) (Prasad, 2011) it uses XHTMLHttpRequests (http://www.w3.org/TR/XMLHttpRequest) to communicate with the back-end. GWT makes it easy to create XHTMLHttpRequests from the front-end. Business objects and data can be serialized to XML, or JSON-format, and transmitted to the front-end where they are de-serialized and further processed. GWAPP uses JSON-format (https://www.json.org) as it is faster to parse, and has less overhead, compared to XML. The back-end uses built-in json parser from *CherryPy* to serialize and de-serialize objects and on the front-end the JSON/XML GWT library *Piriti* (http://code.google.com/p/piriti/) is used. Lastly, automatic gzip compression is activated on the *CherryPy* server, which reduces the size of transmitted data by a third, limiting the bandwidth requirements.